# An Investment Prioritization Model for Wildfire Risk Mitigation Through Power Line Undergrounding

*Saeed Nematshahi[1], Amin Khodaei[1*], Ali Arabnya[1,2]*

[1]Department of Electrical & Computer Engineering, University of Denver, Denver, CO 80210, USA
[2]Quanta Technology, Raleigh, NC 27607, USA
*amin.khodaei@du.edu

## Abstract

Grid-ignited wildfires are one of the most destructive catastrophic events, profoundly affecting the built and natural environments. Burying power lines is an effective solution for mitigating the risk of wildfire ignition. However, it is a costly capital expenditure (CapEx) requiring meticulous planning and investment prioritization. This paper proposes a systematic approach to estimate the potential wildfire ignition damage associated with each transmission line and accordingly offers a priority list for undergrounding. The proposed approach allows electric utilities to make risk-informed decisions for grid modernization and resiliency improvement against wildfires. As a case study, we examine the likelihood of wildfire ignition for each line segment, i.e., between two high-voltage towers, under diverse weather conditions throughout the year. The studies on the standard IEEE 30-bus test system, simulated on 43,712 scenarios, demonstrate the effectiveness of the proposed approach.

## 1. Introduction

In recent decades, climate change has triggered a substantial rise in fire weather conditions around the world [1]. Prolonged droughts and high temperatures further make wildlands more susceptible to spreading fire. Any ignition in these areas can potentially escalate into a large-scale wildfire. Wildfire statistics compiled by the National Interagency Fire Center substantiate a concerning escalation in burned wildlands—from an average of 3.6 million acres per year in 1991-2000 to 6.4 million acres in 2001-2010, and further escalating to 7.5 million acres in 2011-2020 [2]. In the U.S., human activities are responsible for about 9 out of 10 wildfires [3]. These activities include fireworks, campfires, smoking, debris burning, arson, and incidents related to power lines, which are the focus of our research. Fig. 1 depicts the frequency of grid-ignited wildfires in the U.S. from 1992 to 2020 [4].

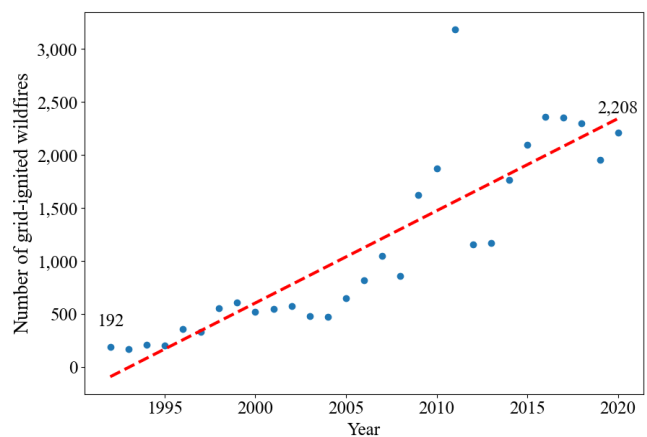

Fig 1. Number of grid-ignited wildfires across the U.S. from 2008 to 2022.

It is evident that the number of wildfires caused by the power grid has increased by more than 11 times, indicating a pressing concern. In terms of intensity, grid-ignited wildfires stand out as some of the most devastating, often holding electric utilities accountable for the damages incurred [5]. Although grid-ignited wildfires are responsible for only 1.43% of all wildfires in the U.S. over the past 3 decades [6], they have been the cause of 7 out of 10 costliest wildfires in U.S. history. In terms of 2024 dollar value, the financial loss for these seven wildfires adds up to over $41.6 billion [7]. This further calls for urgent actions in reducing, or ideally, preventing, grid-ignited wildfires.

An effective strategy for reducing the risk of these formidable wildfires is through burying transmission lines. However, practical deployments have shown that this is a costly investment, reaching $3.3M per mile in some cases [8][9]. The concept of undergrounding power lines is being considered for various reasons. In a study in reference [10], a framework was proposed for designing resilient transmission systems to withstand extreme winds. The authors used a stochastic robust optimization model for a cost-effective undergrounding of power lines with the goal of minimizing load shedding. In a related study, [11] conducted a comparison of the expenses and potential hazards associated with underground and overhead power lines to inform resilient power system planning in the event of windstorms. Authors in [12] suggest improving the resilience of the power grid by replacing certain power lines with natural gas systems. They developed a two-stage robust model for the integrated expansion planning of electricity and gas systems, which considers resilience constraints and uncertainties associated with extreme events.

Existing literature on mitigating grid-ignited wildfires through power line undergrounding includes the studies in [13], which introduce an optimization model utilizing the predefined Wildland Fire Potential Index (WFPI) to select lines for undergrounding, as well as the methodology in [14] that evaluates the susceptibility of the transmission grid to grid-ignited wildfires in terms of their impact on other lines and the potential to cause power outages. The former approach simplified wildfire risk assessment by overlaying the WFPI with grid topology without running specific wildfire simulations, and the latter focused solely on the impact of wildfires on power grid outages, without considering the financial consequences. Aiming at addressing the gap, the main contributions of our paper are twofold. First, it simulates grid-ignited wildfires and quantifies the resulting damage costs. Second, it identifies the optimal set of lines for undergrounding to minimize financial losses due to grid-ignited wildfires. We present a systematic approach integrating landscape, weather, and power system data to estimate the total damage to the environment and other lines in the event of a grid-ignited wildfire. We accordingly calculate the monetary value representing the potential damage (financial risk exposure), which will be used as input to the undergrounding investment prioritization process.

The rest of the paper is organized as follows. Section 2 provides a description of the proposed methodology, detailing the theoretical framework and algorithmic steps. Section 3 presents a comprehensive case study to demonstrate the methodology's practical application, including numerical results and discussions on the findings. Section 4 offers a conclusion, summarizing key insights and main contributions.

## 2. Proposed Methodology

Our proposed methodology for prioritizing transmission line undergrounding involves five integral steps: data integration, wildfire simulation, financial loss assessment, likelihood analysis, and the establishment of a priority list.

**Step 1:** A geographical map delineating the transmission system is developed. Ignition points are strategically positioned along each power line, necessitating precise identification by their longitude and latitude coordinates. For each transmission line, the number of ignition points is equal to the number of transmission towers along the line plus one. This ensures that there is one ignition point between each pair of adjacent towers. Additionally, topographical maps and meteorological data pertinent to the study area within the designated time frame are required. The topographical data can be sourced from LandFire, a Landscape Fire and Resources Management Planning tool, as depicted in Fig. 2. Meteorological data can be obtained from the National Solar Radiation Database (NSRDB) [15]. Next, the timeframe (season) weather data including the ambient temperature, humidity, wind speed, and wind direction scenario are set, and all data is subsequently inputted into FARSITE, a fire spread simulator developed by the U.S. Forest Service [16].

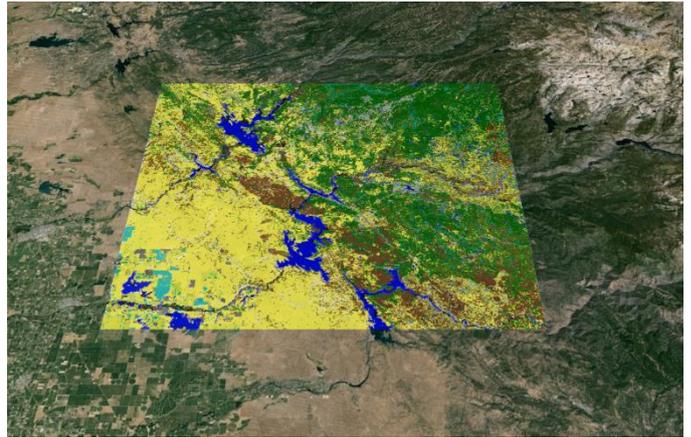

Fig. 2. A topographical map, including eight layers of elevation, slope, aspect, fuel model, canopy cover, stand height, canopy base height, and canopy bulk density.

**Step 2:** It involves analyzing the wildfire behavior for each scenario, taking into account the investigation resolution and suppression assumptions. To analyze wildfire behavior, we have used FARSITE which predicts the spread and intensity of wildfires over time, considering various factors such as weather, fuel, and topography. The inputs to FARSITE include ignition files, weather data streams, burned period, and fuel moisture files. This encompasses various landscape parameters, including elevation, slope, aspect, fuel models, canopy cover, stand height, canopy base height, canopy bulk density, and physical barriers. Additionally, the weather data comprises temperature, humidity, wind direction, and wind speed. FARSITE processes these inputs and produces outputs such as arrival time, spread direction, and rate of spread. Using these output files, the corresponding burned area for each wildfire scenario is determined. Given that grid-ignited wildfires are classified as single-ignition fires, FARSITE is specifically tailored to model these types of wildfires. This configuration ensures that the simulation accurately reflects the behavior and spread patterns of fires originating from a single ignition point, thereby providing more precise and reliable simulations. The output of this step is a fire status matrix for the wildfire scenario, where each cell indicates the fire status (1 for the presence of fire and 0 for no fire) within a defined resolution unit. The number of cells depends on the area of study and the desired resolution.

**Step 3:** The burned area is superimposed on the landscape and the grid map to assess its impact on the surroundings and the grid itself, respectively. The affected area and power line outages are identified and recorded. Next, the average financial loss for the burned area and the average cost for transmission line reconstruction are determined. Afterward, the affected environment and affected lines, along with the average cost for each, are sent to financial loss assessment. The total cost of damage caused by wildfires originated by each specific power line is calculated. This financial loss encompasses several critical components, including liability to third parties (covering property damage, personal injury, inverse condemnation liabilities, and other related claims from affected individuals and businesses) and direct damage to the transmission grid (entailing costly repairs or replacement of damaged infrastructure, loss of service, and restoration expenses).

**Step 4:** Outputs from different scenarios are combined using a weighted average approach. To enhance the robustness of our study in the face of unforeseen weather patterns, it is essential to adjust the historical data, allowing our methodology to better account for and adapt to future conditions. Thus, the probabilities for different seasons and wind direction shifts are used to calculate the final weighted damage cost.

**Step 5:** The transmission lines are ranked according to the weighted average cost of damage calculated for each line. This process results in a sorted list of power lines, enabling the utility company to prioritize its existing investments.

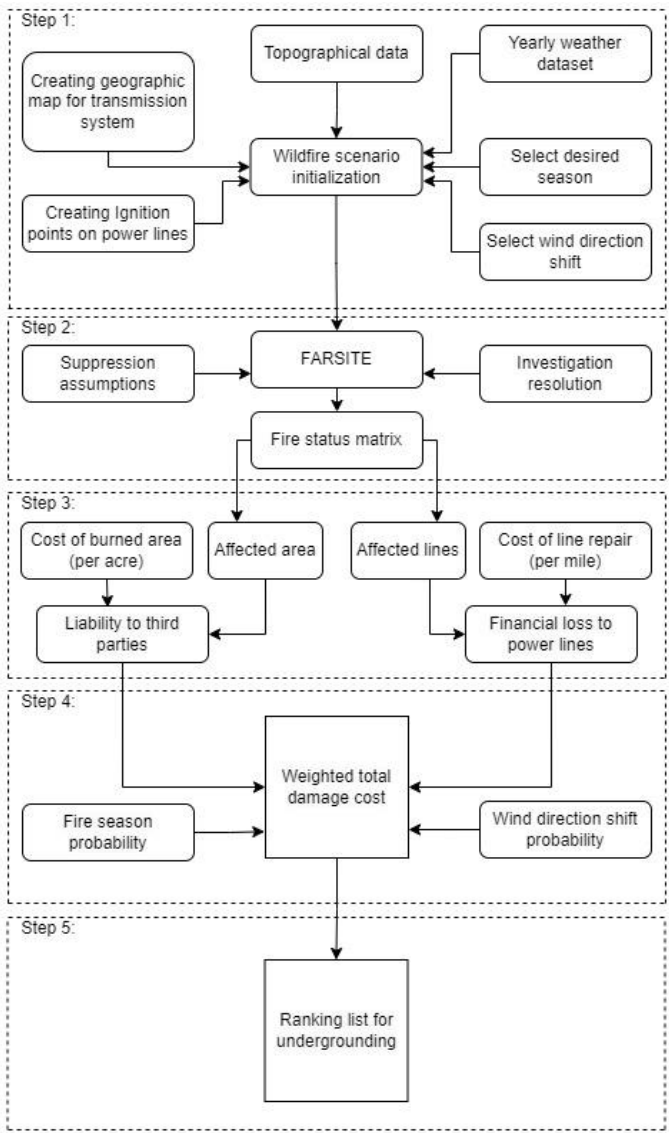

Fig. 3. The schematic of the proposed methodology for undergrounding investment prioritization strategy.

Fig. 3 illustrates the overall structure of the proposed methodology. This assessment allows electric utilities to allocate their limited budget effectively, focusing on undergrounding the lines with the highest potential cost of damage. To ensure the accuracy of our findings, we reassessed the wildfire scenarios for various wind directions, each with a deviation from the others and a reasonable probability. To conduct a more comprehensive analysis, we included historical meteorological data from all four seasons for the studied area, considering the likelihood of wildfire occurrences. This approach significantly broadens the spectrum of potential wildfire scenarios, and consequently, the ranking list would be more reliable.

## 3. Simulations

### 3.1. Case Study

The IEEE 30-bus test system is used, as referenced in [17], for the case study in this paper. Fig. 4 illustrates the transmission system, which comprises 41 branches, including seven links and 34 transmission lines. Notably, each power line has the potential to cause wildfires.

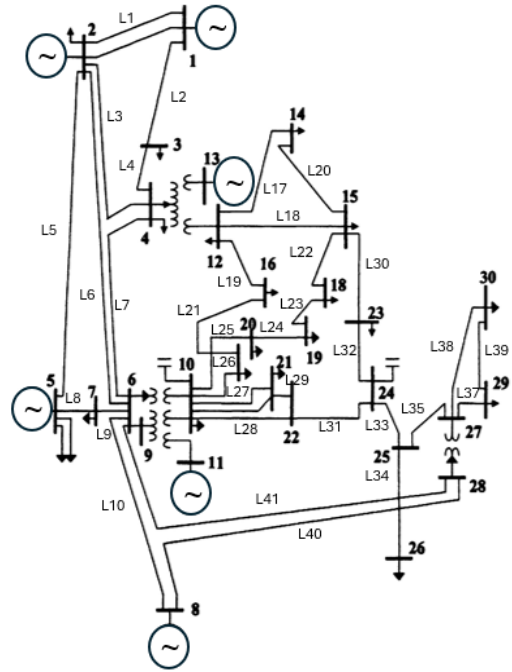

Fig. 4. Single line diagram of IEEE 30-bus test system.

The scenarios are defined based on the following 3 assumptions:

1) To consider the possibility of a single ignition between each pair of adjacent towers in a transmission line, we need *N+1* ignition points, where *N* is the number of towers on that line. It is assumed that the distance between two high-voltage towers is 0.3 miles. Thus, the number of ignition points for each transmission line is proportional to the length of the line, resulting in a total of 1366 ignition points across the transmission grid. Table 1 shows the length of each line and the corresponding number of potential ignition points.

2) The 2022 weather data was modified for wind direction to create eight distinct scenarios, each differing by 45 degrees from the next while maintaining a reasonable probability. The probabilities for each are presented in Table 2.

Table 1. Transmission lines Data

| Branch ID | Length (Mile) | Ignition points | Branch ID | Length (Mile) | Ignition points |
|---|---|---|---|---|---|
| 1 | 11.3 | 37 | 24 | 4.44 | 13 |
| 2 | 15.2 | 49 | 25 | 15.1 | 47 |
| 3 | 28.4 | 91 | 26 | 9.17 | 29 |
| 4 | 4.8 | 15 | 27 | 6.39 | 21 |
| 5 | 31.3 | 99 | 28 | 9.59 | 31 |
| 6 | 33.6 | 107 | 29 | 4.05 | 13 |
| 7 | 16.3 | 51 | 30 | 14 | 45 |
| 8 | 10.9 | 35 | 31 | 14 | 45 |
| 9 | 9.5 | 29 | 32 | 8.79 | 27 |
| 10 | 12.8 | 41 | 33 | 7.55 | 23 |
| 17 | 14.2 | 45 | 34 | 9.58 | 29 |
| 18 | 9.79 | 31 | 35 | 5.76 | 19 |
| 19 | 13.6 | 43 | 37 | 6.61 | 21 |
| 20 | 7.64 | 25 | 38 | 14.5 | 45 |
| 21 | 5.45 | 17 | 39 | 7.94 | 25 |
| 22 | 6.67 | 21 | 40 | 28.6 | 89 |
| 23 | 5.87 | 19 | 41 | 28.4 | 89 |

Table 2. Wind direction shift probability

| 0° | 45° | 90° | 135° | 180° | 225° | 270° | 315° |
|---|---|---|---|---|---|---|---|
| 30% | 15% | 10% | 7.5% | 5% | 7.5% | 10% | 15% |

3) Given that grid-ignited wildfires can occur throughout the year, it is essential to account for all seasons and extend the study's duration to at least one full year. Since the probability of ignition varies by season, the historical weather data for each of the four seasons has been considered based on the likelihood of wildfire occurrence, as shown in Table 3. Wildfire probability fluctuates with the seasons in wildfire-prone areas. In spring, the likelihood is approximately 20% due to rising temperatures and lingering dry vegetation. During summer, it increases to around 40% due to high temperatures and dry weather. Fall sees a probability of about 30% owing to sustained dry conditions and strong winds, while winter brings the probability down to 10% due to cooler temperatures and increased moisture levels. However, it's important to note that these percentages can vary significantly based on specific regional and annual conditions.

Table 3. Fire season probability

| Winter | Spring | Summer | Fall |
|---|---|---|---|
| 10% | 20% | 40% | 30% |

Based on the above inputs, a total of 43,712 wildfire ignition scenarios are generated. This number includes 1,366 grid-ignited wildfire scenarios, each modeled with 32 distinct weather streams. Among these 32 weather streams, the actual summer data, with no changes in wind direction, exhibits the highest probability of occurrence at 0.12. Conversely, the winter scenario with a 180-degree shift in wind direction has the lowest probability of occurrence at 0.005.

*3.2. Numerical Results*

The experiments were conducted on a system with a 13th Gen Intel(R) Core (TM) i7-13700 processor running at 2.10 GHz and 16 GB of DDR4 RAM. The total execution time for the code was about 85 hours.

In the proposed methodology, the total damage cost of a grid-ignited wildfire includes damage to third parties and the power grid itself.

Firstly, we need to calculate the cost of damage to third parties. This can be extracted from the burned area matrix, an outcome of FARSITE, where each cell represents the status of the fire for a 0.075-mile by 0.075-mile area of study. Our study area consists of a 517 by 464 matrix. In this matrix, the fire status is represented as 1 if it reaches area specific cell; otherwise, it remains 0.

Next, the number of cells with the burned status for each scenario is identified, and the total burned area is calculated by adding up all the burned cells. This process will be carried out separately for each ignition point on every transmission line. The average amount identified as the damage to third parties for that line. Afterward, the cost of damage to third parties for a specific area will be calculated by multiplying the acres of the burned area by $20k, which is a medium-range financial loss based on the findings in [18]. This amount may increase significantly if a wildfire approaches an urban area, resulting in a higher number of destroyed structures.

Secondly, we need to calculate the cost of damage to the power grid itself by overlaying the burned area with the transmission grid map. After determining the total miles of transmission lines necessitating restoration, we assumed a $200k per mile for replacing the burned transmission lines.

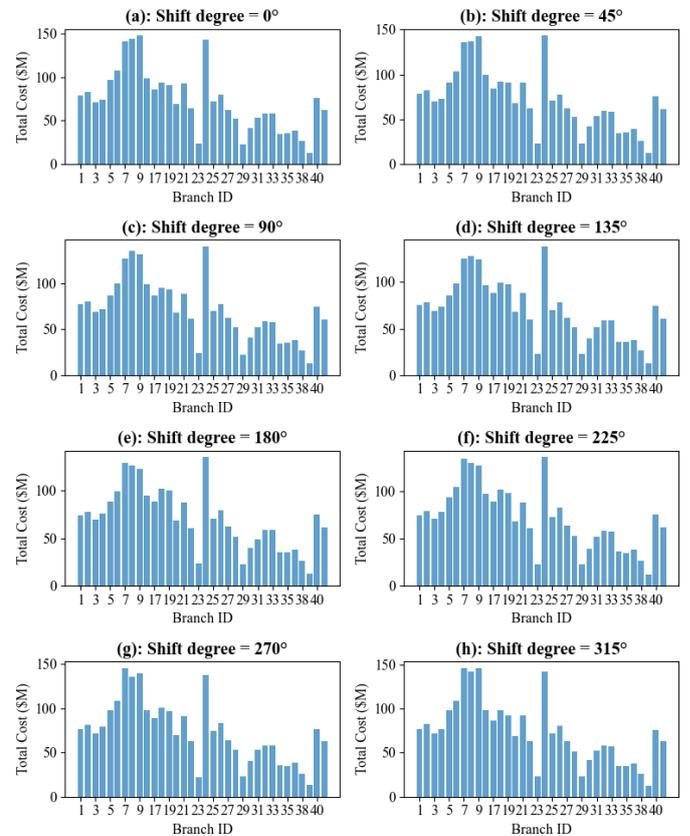

Fig. 5. Wind direction impact on wildfire total cost.

By adding these two amounts, we have the total damage cost for each line. To ensure more reliable study, we need to assess the total costs for eight different wind directions with reasonable probability. Fig. 5 illustrates the impact of the wind direction on the total financial loss caused by a grid-ignited wildfire. Each plot shows the total cost for 34 transmission lines.

As shown, changes in wind direction may exacerbate or alleviate the total financial loss. Fig. 5a shows the total cost for each line with no change in the wind direction, while other subsections show a shift in degree from the actual data. This shift leads to a change in wildfire spread and the corresponding burned area. Thus, the total cost is affected.

Considering all possibilities, a reasonable average will be calculated based on the probabilities expressed in Table 2.

In addition, a period of at least one year consisting of different seasons is needed to have a wide range of temperatures, humidity, and wind speed. Fig. 6 illustrates the changes in total financial loss over four seasons in 2022.

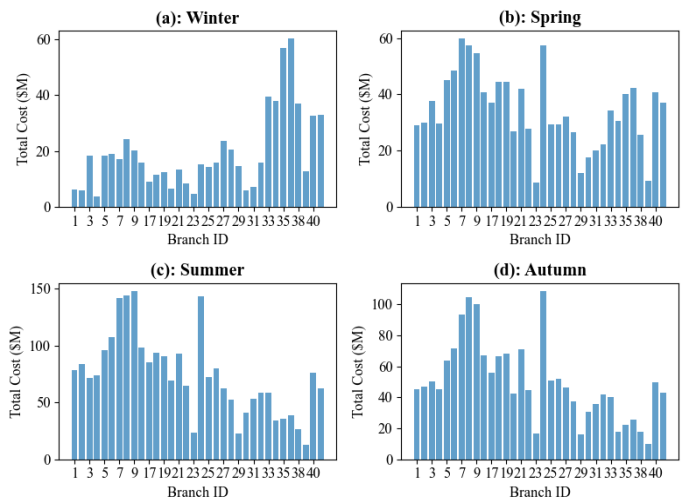

Fig. 6. Seasonal weather impact on wildfire total cost.

It is worth noting that Fig. 5c exhibits the highest wildfire costs compared to others. This observation clearly demonstrates the influence of high temperatures and low humidity on wildfire propagation in the summer. Furthermore, the ranking of the costliest line varies throughout the year, requiring a year-long investigation. In winter, line 37 poses the highest risk, while in spring, line 7 is the costliest. During the summer, line 9 ranks first in potential damage, and in Fall, line 24 surpasses others.

Since the likelihood of wildfire ignition is not the same during all seasons, without loss of generality, we assume a set of probabilities, as shown in Table 3.

Considering a set of eight wind directions and four seasons, the financial cost caused by a grid-ignited wildfire starting from each line is estimated. As an example, for line 24, we have 32 values, each representing a specific season with a specific modification in wind direction. Each value is calculated as the average total damage of all 13 single-ignition wildfire scenarios ignited from line 24. The model outcome is represented in Table 4.

Table 4. Total financial damage for line 24 (in Million Dollars)

|        | 0°  | 45° | 90° | 135° | 180° | 225° | 270° | 315° |
|--------|-----|-----|-----|------|------|------|------|------|
| Winter | 15  | 15  | 16  | 16   | 15   | 15   | 15   | 15   |
| Spring | 58  | 57  | 56  | 56   | 56   | 58   | 58   | 58   |
| Summer | 143 | 143 | 141 | 138  | 135  | 136  | 137  | 142  |
| Fall   | 108 | 109 | 106 | 102  | 100  | 103  | 105  | 107  |

Integrating Table 2 and Table 3, the weight matrix will be defined in Table 5. It is clearly seen that the maximum weight is observed in the summer season with no shift in wind direction, keeping it as it is in the actual weather data.

Table 5. Weight matrix

|        | 0°   | 45°   | 90°  | 135°   | 180°  | 225°   | 270° | 315°  |
|--------|------|-------|------|--------|-------|--------|------|-------|
| Winter | 0.03 | 0.015 | 0.01 | 0.0075 | 0.005 | 0.0075 | 0.01 | 0.015 |
| Spring | 0.06 | 0.03  | 0.02 | 0.015  | 0.01  | 0.015  | 0.02 | 0.03  |
| Summer | 0.12 | 0.06  | 0.04 | 0.03   | 0.02  | 0.03   | 0.04 | 0.06  |
| Fall   | 0.09 | 0.045 | 0.03 | 0.0225 | 0.015 | 0.0225 | 0.03 | 0.045 |

Finally, the weighted average value for the entire year is calculated using Table 4 and Table 5, amounting to $101.31M for line 24.

By applying the same process to each transmission line, we can compare different lines in Fig. 7.

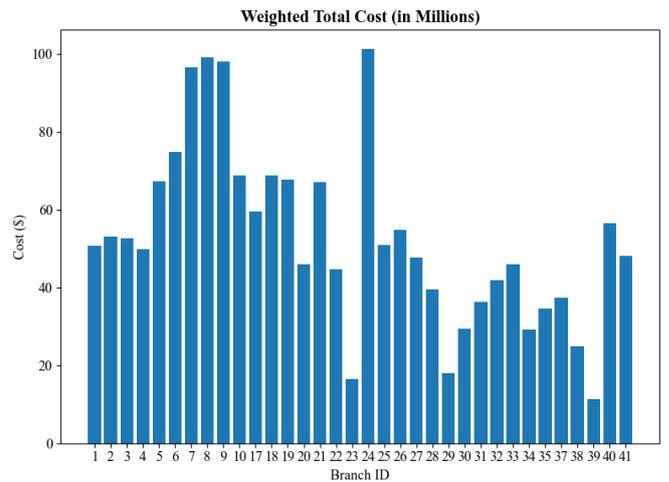

Fig. 7. Comparison of total costs across different transmission lines.

As illustrated in Fig. 7, line 24 ranks first in the total damage cost of grid-ignited wildfires, followed by lines 8, 9, and 7. The high cost associated with line 24 can be attributed to its location, which is both in a fire-prone area and in close proximity to other lines. Additionally, it is important to note that line 24 is 4.44 miles long, making it the second shortest line. This study demonstrates that with strategic prioritization, replacing just 4.44 miles of line 24 can achieve greater risk mitigation than replacing longer lines, such as line 3, which is 28.4 miles long.

The prioritization of investment in undergrounding aims to minimize the potential financial loss caused by the power grid.

Thus, the ranking list for undergrounding is provided in Table 6.

Table 6. Ranking list for undergrounding

| Ranking List | Branch ID | Ranking List | Branch ID |
|---|---|---|---|
| 1 | 24 | 18 | 4 |
| 2 | 8 | 19 | 41 |
| 3 | 9 | 20 | 27 |
| 4 | 7 | 21 | 33 |
| 5 | 6 | 22 | 20 |
| 6 | 18 | 23 | 22 |
| 7 | 10 | 24 | 32 |
| 8 | 19 | 25 | 28 |
| 9 | 5 | 26 | 37 |
| 10 | 21 | 27 | 31 |
| 11 | 17 | 28 | 35 |
| 12 | 40 | 29 | 30 |
| 13 | 26 | 30 | 34 |
| 14 | 2 | 31 | 38 |
| 15 | 3 | 32 | 29 |
| 16 | 25 | 33 | 23 |
| 17 | 1 | 34 | 39 |

Leveraging the list of priorities, utilities can allocate their limited budget strategically to underground power lines. In our case study, the limited budget is allocated to lines 24, 8, 9, etc., respectively.

## 4. Conclusion

In this paper, we proposed a practical methodology for calculating the total cost of a potential wildfire ignited by transmission lines. This model provides electric utilities with a practical risk-informed strategy for undergrounding their transmission assets, prioritizing those with the highest potential risk of damage to third parties and power infrastructure. A ranking list of transmission lines is obtained, and based on that, an electric utility can allocate its limited budget to underground its power lines. Historical meteorological data for four seasons were extracted with a designed probability and modified for several wind directions to achieve more calibrated results. The numerical results demonstrate the effectiveness of the proposed investment prioritization model for wildfire mitigation through undergrounding.

## 5. References

[1] Jones, M. W., Abatzoglou, J. T., Veraverbeke, S., Andela, N., Lasslop, G., Forkel, M., et al. (2022). Global and regional trends and drivers of fire under climate change. Reviews of Geophysics, 60, e2020RG000726.
[2] National Interagency Coordination Center, Source available: https://www.nifc.gov/fire-information/statistics/wildfires.
[3] J. K. Balch, B. A. Bradley, J. T. Abatzoglou, R. C. Nagy, E. J. Fusco, A. L. Mahood. (2017). Human-started wildfires expand the fire niche across the United States. Proceedings of the National Academy of Sciences 114, no. 11: 2946-2951.
[4] K. C. Short, "Spatial Wildfire Occurrence Data for the United States, 1992-2020 [FPA_FOD_20221014]," 6th ed., Forest Service Research Data Archive, Fort Collins, CO, 2022. [Online]. Available: https://doi.org/10.2737/RDS-2013-0009.6 [Accessed June 2024].
[5] A. Arab, A. Khodaei, R. Eskandarpour, M. P. Thompson and Y. Wei, "Three Lines of Defense for Wildfire Risk Management in Electric Power Grids: A Review," in IEEE Access, vol. 9, pp. 61577-61593, 2021.
[6] K. C. Short, "Spatial Wildfire Occurrence Data for the United States, 1992-2020 [FPA_FOD_20221014]," 6th ed., Forest Service Research Data Archive, Fort Collins, CO, 2022. [Online]. Available: https://doi.org/10.2737/RDS-2013-0009.6 [Accessed April 2024].
[7] The National Fire Protection Association (NFPA). Wildland Fire Statistics. [Online]. Available: https://www.nfpa.org/Education-and-Research/Research/NFPA-Research/Fire-Statistical-reports/Wildland-Fire-Statistics [Accessed April 2024].
[8] M. Nauman, "Crews Safely Complete Construction, Energization of 350 More Miles of Underground Powerlines in 2023". Pacific Gas and Electric Company. [Online]. Available: https://www.pgecurrents.com/articles/3900-crews-safely-complete-construction-energization-350-miles-underground-powerlines-2023. [Accessed July 2024].
[9] Pacific Gas and Electric Company. 2023 General Rate Case. [Online]. Available: https://www.pge.com/en/regulation/general-rate-case.html [Accessed April 2024].
[10] D. N. Trakas and N. D. Hatziargyriou, "Strengthening Transmission System Resilience Against Extreme Weather Events by Undergrounding Selected Lines," in IEEE Transactions on Power Systems, vol. 37, no. 4, pp. 2808-2820, July 2022.
[11] L. Souto and S. Santoso, "Overhead versus Underground: Designing Power Lines for Resilient, Cost-Effective Distribution Networks under Windstorms," 2020 Resilience Week (RWS), Salt Lake City, UT, USA, 2020, pp. 113-118.
[12] C. Shao, M. Shahidehpour, X. Wang, X. Wang and B. Wang, "Integrated Planning of Electricity and Natural Gas Transportation Systems for Enhancing the Power Grid Resilience," in IEEE Transactions on Power Systems, vol. 32, no. 6, pp. 4418-4429, Nov. 2017.
[13] S. Taylor, and L.A. Roald. "A framework for risk assessment and optimal line upgrade selection to mitigate wildfire risk." Electric Power Systems Research 213 (2022): 108592.
[14] B. Sohrabi, A. Arabnya, M. P. Thompson and A. Khodaei, "A Wildfire Progression Simulation and Risk-Rating Methodology for Power Grid Infrastructure," in IEEE Access, doi: 10.1109/ACCESS.2024.3439724.
[15] National Solar Radiation Database. [Online]. Available: https://nsrdb.nrel.gov/. [Accessed: June 19, 2024].
[16] Finney, Mark A. 1998. FARSITE: Fire Area Simulator-model development and evaluation. Res. Pap. RMRS-RP-4, Revised 2004, Ogden, UT: U.S. Department of Agriculture, Forest Service, Rocky Mountain Research Station. 47 p.
[17] S. Nematshahi, A. Khodaei, A. Arabnya, "Risk assessment of transmission lines against grid-ignited wildfires," 2025 IEEE PES Grid Edge Technologies Conference & Exposition (Grid Edge), San Diego, CA, USA, 2025, pp. 1-5, in press.
[18] P. Howard, "Flammable planet: Wildfires and the social cost of carbon." New York: NYU Law School Institute for Policy Integrity (2014).